\documentclass[a4paper]{article}
\usepackage{ASVspoof}
\usepackage{epsfig,amssymb,amsmath}
\usepackage{hyperref}
\usepackage{siunitx}

\usepackage{epsfig,amssymb,amsmath}
\usepackage{mathtools}

\usepackage{multirow}
\usepackage{stfloats}
\usepackage{float}
\usepackage{amsmath,graphicx,booktabs,amsfonts,amssymb,utfsym,multirow,makecell,url,color, colortbl,subcaption,siunitx}
\usepackage{soul}

\title{Exploring WavLM Back-ends for Speech Spoofing and Deepfake Detection}

\name{Theophile Stourbe, Victor Miara, Theo Lepage, Reda Dehak}

\address{ EPITA Research Laboratory (LRE), France \\
    {\small \tt \{theophile.stourbe, victor.miara, theo.lepage, reda.dehak\}@epita.fr}
}

\begin{document}

\maketitle

\begin{abstract}

This paper describes our submitted systems to the ASVspoof 5 Challenge Track 1: Speech Deepfake Detection - Open Condition, which consists of a stand-alone speech deepfake (bonafide vs spoof) detection task. Recently, large-scale self-supervised models become a standard in Automatic Speech Recognition (ASR) and other speech processing tasks. Thus, we leverage a pre-trained WavLM as a front-end model and pool its representations with different back-end techniques. The complete framework is fine-tuned using only the trained dataset of the challenge, similar to the close condition. Besides, we adopt data-augmentation by adding noise and reverberation using MUSAN noise and RIR datasets. We also experiment with codec augmentations to increase the performance of our method. Ultimately, we use the Bosaris toolkit for score calibration and system fusion to get better Cllr scores. Our fused system achieves $0.0937$ minDCF, $3.42\%$ EER, $0.1927$ Cllr, and $0.1375$ actDCF.

\end{abstract}

\newcommand\blfootnote[1]{%
  \begingroup
  \renewcommand\thefootnote{}\footnote{#1}%
  \addtocounter{footnote}{-1}%
  \endgroup
}


\section{Introduction}
With the development of Deep Neural Networks (DNN), speech synthesis and Voice Conversion (VC) are making significant progress in generating natural speech audio. This increases the relevance of spoofing speech detection to protect speaker identities and make biometric systems based on Automatic Speaker Verification (ASV) systems more robust. The challenge of anti-spoofing speech recognition is to detect the artifacts produced by the generation process or the VC of the spoof attacks. Over the past few years, many challenges have been proposed to promote the consideration of spoofing and speech deepfakes detection. ASVspoof is a series of challenges that focus on developing and benchmarking systems to detect and mitigate various spoofing attacks in ASV systems. ASVspoof 5 \cite{ASVS5,ASVSpoof5evalplan}, the fifth edition in this series, comprises two different tracks. The first track, which is the goal of our work, is to develop a countermeasure system to detect deepfake audio. This task comprises two challenges; the first one (the close condition) consists of creating a system using only the training dataset of the challenge. In the other challenge, the open condition, pre-trained models, and other training datasets are allowed, except those used to generate the test data.

Initially, most anti-spoofing methods were based on DNN processing frame-level handcrafted acoustic features, such as MFCC, LFCC, log-linear filter bank, CQCCs, and CQT spectrogram \cite{WU2015130}. Other approaches adopt raw waveforms as the input to train an end-to-end DNN to detect deepfake audio \cite{tak2021Rawnet2}. 

With the success of large self-supervised models for speech processing, many solutions were proposed using HuBERT \cite{Hsu_2021}, wav2vec 2.0 \cite{baevski_wav2vec_2020} and XLS-R \cite{babu22_interspeech} models as a feature extractor for a downstream system. Wang et Yamagishi \cite{wang22_odyssey} explore the use of HuBERT and wav2vec2.0 as a feature extractor for audio spoofing detection. Tak \emph{et al.} \cite{tak22_odyssey} succeed in using XLS-R as a front-end speech features extractor for an AASIST model and get a significant improvement over previous techniques. These two works show that fine-tuning the front-end module is necessary to get better performances since this model was only trained with bonafide data. We can conclude that using large self-supervised pre-trained models for spoofing detection is very efficient. Thus, we decided to develop a submission for ASVspoof 5 based on the WavLM model \cite{chen2022WavLM} which demonstrates impressive results in many areas related to speech processing, emotion recognition \cite{diatlova24_odyssey} and speaker recognition \cite{peng2023MHFA, miara2024ssl}.

Our system is based on the WavLM Base model as a front-end feature extractor; pre-trained on $960h$ of Librispeech \cite{panayotov2015librispeech} which is compliant with the ASVspoof~5 open condition track rules\cite{ASVS5}. We used the CNN encoder layer and the first 12-th Transformers encoder layers as features for a downstream back-end system. We use two different back-ends to aggregate all representations into one embedding vector: Weighted Average (WA) pooling and Multi-Head Factorized Attentive (MHFA) pooling proposed for speaker recognition in \cite{peng2023MHFA}. Several works show that data-augmentation is necessary to learn robust detection systems and to avoid over-fitting and improve generalization \cite{COHEN202256, Ge2024}. Codec augmentation, background noise augmentation, reverberation as well as RawBoost \cite{tak2022rawboost} were tested and used to improve the generalization and robustness of our approach.

The paper is organized as follows: Section \ref{sec:datasets} presents the dataset used in model training, model scoring, and systems fusion and calibration steps. Section \ref{sec:augmentation} details our approach for data augmentation, and Section \ref{sec:model} describes the front-end and the back-end of the models we developed. Section \ref{sec:setup} describes the different hyper-parameters used for the training process. The results and performances are presented in Section \ref{sec:result}. The conclusions are given in Section \ref{sec:conclusion}. 

\begin{figure*}[t]
    \centering
    \begin{subfigure}[T]{0.421\textwidth}
        \centering
        \includegraphics[width=0.93\linewidth]{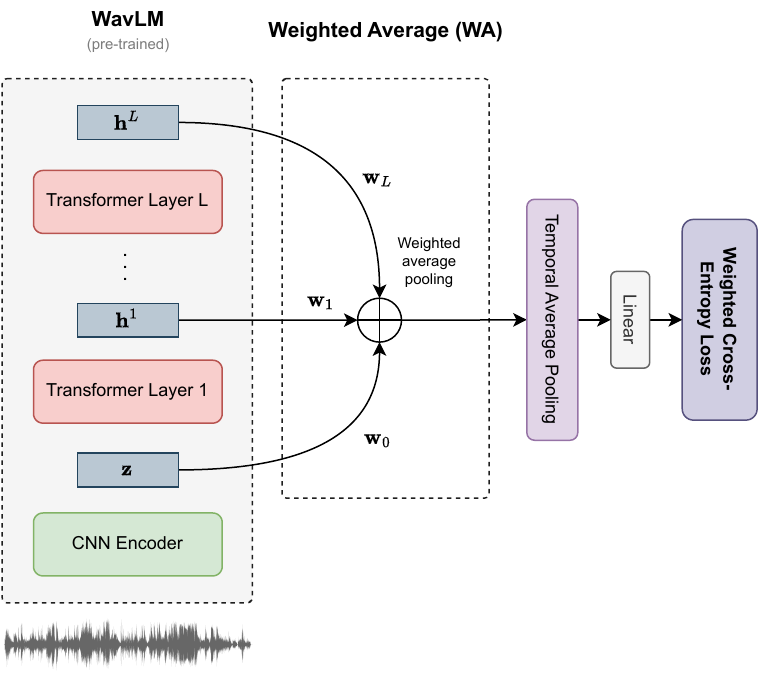}
        \caption{Weighted Average (WA)}
    \end{subfigure}%
    \begin{subfigure}[T]{0.579\textwidth}
        \centering
        \includegraphics[width=0.93\linewidth]{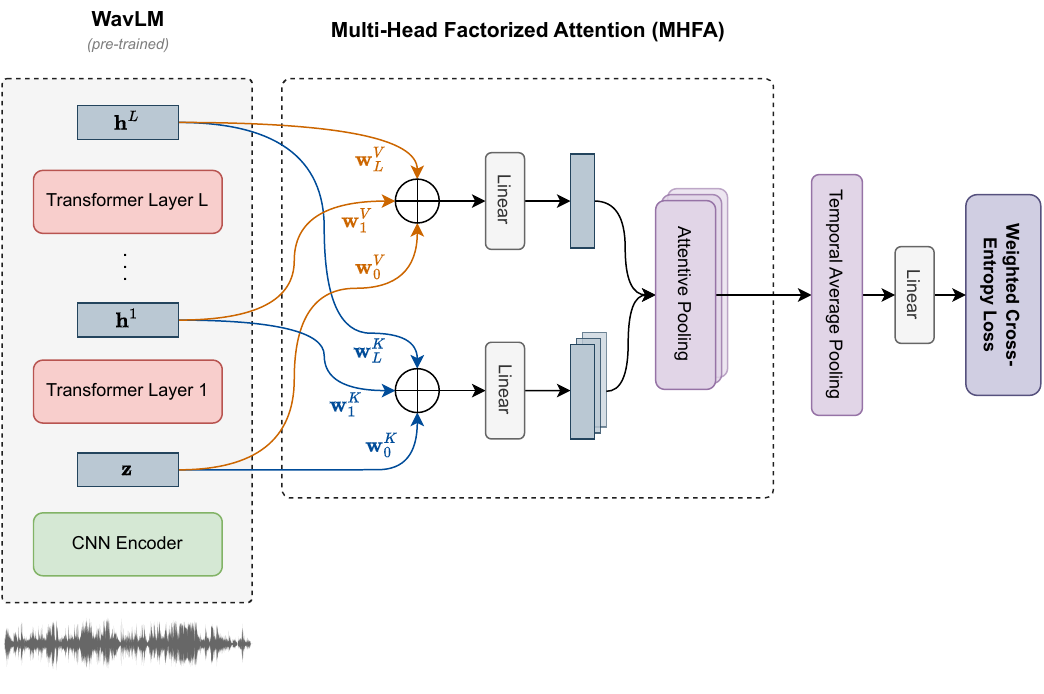}
        \caption{Multi-Head Factorized Attention (MHFA)}
    \end{subfigure}
    \caption{Diagram of our framework for fine-tuning WavLM with WA (a) and MHFA (b) back-ends.}
    \label{fig:training_framework}
\end{figure*}

\sisetup{
    output-decimal-marker = {,}
}

\section{Datasets}
\label{sec:datasets}
\begin{table}[hbt]
    \centering
    \caption{Summary of the contents of the training dataset and development subsets.}
    \label{tab:dataset}
\begin{footnotesize}
    \begin{tabular}{llSS}
    \toprule
\multirow{2}{*}{\textbf{Subset}}& \multirow{ 2}{*}{\textbf{Usage}}  & \multicolumn{2}{c}{\textbf{\# utterances}} \\
          &       & \text{Bonafide} & \text{Spoof} \\
    \midrule
    Training & Model training & 18,797 & 163,560 \\
    Development (1) & Scoring & 8,245 & 28,846 \\
    Development (2) & Fusion and calibration & 23,089 & 80,770 \\
    \bottomrule
    \end{tabular}
\end{footnotesize}
\end{table}
For system development, ASVspoof released a training dataset containing $182,357$ utterances and a development dataset containing $140,950$ utterances distributed as $16$ kHz, $16$ bits per sample FLAC files. We used the training dataset to fine-tune the WavLM and the back-end models. The development dataset was split into two parts: the \textbf{scoring} set used to compute and compare the models’ performances; and the \textbf{fusion and calibration} set used for score calibration and model fusion. The contents of each dataset are summarized in Table \ref{tab:dataset}.

\section{Data-augmentation}
\label{sec:augmentation}

Previous works show the importance of data-augmentation to obtain better performance and improve generalization. During the model training, we tested a combination of 4 different data-augmentation techniques described below.
\begin{itemize}
    \item Utterances are augmented with background noises randomly selected from the 929 various noises of the MUSAN corpus \cite{snyder2015musan}. The SNR is uniformly sampled between 0 and 15 dB. 
    \item To apply reverberation, we convolve the input audio segment with an impulse response randomly sampled from the Simulated Room Impulse Response Database \cite{ko2017rir}
    \item We use torchaudio library to apply low and high-quality mp3 and ogg encoder. m4a could not be tested as it is not implemented in the torchaudio library. We also tested four trans-codecs configuration: high mp3 $\rightarrow$ high ogg, low mp3 $\rightarrow$ low ogg, high mp3 $\rightarrow$ low ogg, high ogg $\rightarrow$ low mp3.
    \item We also experiment with RawBoost similar to \cite{takrawboost}. This method requires no additional data sources, e.g., noise recordings or impulse responses, and it is data, application, and model-agnostic. This data augmentation algorithm is not used in combination with the other techniques mentioned above.
\end{itemize}

All our data augmentation was implemented online (parallel) in our training framework so that the model is trained on different data at each epoch. For each sample, we first apply or not randomly selected noise from MUSAN noise, random reverberation from RIR dataset, or both. Second, we apply or not a random codec augmentation. In the case of RawBoost, we tested the different algorithms proposed in \cite{takrawboost}.


\section{Models}
\label{sec:model}

\subsection{Baseline: ResNet-based spoof detection system}

We chose a ResNet-based backbone as a baseline system and submitted this system for the closed condition. We chose this model because we get better performance compared to the RawNet2 \cite{tak2021Rawnet2} and the AASIST \cite{jung2022aasist} methods proposed in the ASVspoof 5 Toolkit.

We rely on the Fast ResNet-34 architecture, described in \cite{chung2020resnet}, processing $40$-dimensional log-mel spectrogram input features with a Hamming window of $25$ ms length and $10$ ms frame-shift. The encoder dimension is set to 512 and we add a ReLU activation function followed by a linear layer for the spoofing detection task.

\subsection{WavLM-based spoof detection system}
\label{sec:methods_wavlm}

WavLM \cite{chen2022WavLM} is a Transformer-based model designed for Automatic Speech Recognition (ASR). It is pre-trained in a self-supervised way that also captures non-ASR information. During the pre-training, the model processes raw audio through a multi-layer convolutional feature encoder, transforming a sequence $\left\{\mathbf{x}_t\right\}_{t=1}^T$ of $T$ time windows to produce $\left\{\mathbf{z}_t\right\}_{t=1}^T$. These representations are then subject to noise and overlapping before masking and fed into the Transformer encoder, which outputs a series of hidden states $\left\{\mathbf{h}^l\right\}_{l=1}^L$, where $L$ denotes the number of Transformer layers. Additionally, the model incorporates gated relative position bias, enhancing its ability to focus on relevant speech features. WavLM is trained on a masked speech denoising and prediction task which implicitly models speaker and speech-related information as the objective is to predict the pseudo-labels of the original speech on masked regions.

\subsubsection{Weighted Average (WA) back-end}

Several works show that the intermediate representation of the self-supervised model contains essential features that can be used in various speech downstream tasks. Generally, the top layers, which are closer to the objective of the pre-training task, tend to be the most helpful for automatic speech recognition (ASR). In contrast, the speech and speaker features are mainly represented in the low- and mid-level features, which carry most information about speech signals. Thus, using only the last Transformer layer’s output might be sub-optimal for speech spoofing detection. 

As shown in Figure \ref{fig:training_framework}-a, following \cite{chen2022WavLM}, using the outputs $\mathbf{z}_t$ and $\left\{\mathbf{h}^l_t\right\}_{l=1}^L$ of the $l$-th transformer layer for each frame $t$, we learn a weighted average of all these outputs to generate a new frame representation $O_t$ such that
\begin{equation}
    O_t = w_0 \mathbf{z}_t + \sum_{l=1}^L w_l \mathbf{h}^l_t,
\end{equation}
where $\{w_k \mid 0 \leq k \leq L\}$ represent the learnable weights.

Next, the weighted frame-level representation $O_t$ is fed into a temporal average pooling layer followed by a fully connected layer to obtain the final score for spoof detection.

\subsubsection{Multi-Head Factorized Attention (MHFA) back-end}

Following \cite{peng2023MHFA}, Multi-Head Factorized Attention (MHFA) back-end (Figure \ref{fig:training_framework}-b) consists of aggregating layer-wise outputs from WavLM's transformer layers into an attentive pooling mechanism that clusters frame-level representations into acoustic units discovered by the transformer model. The frame-level representations are then aggregated (pooled) within each cluster and combined to produce the final frame embedding. This mechanism allows frame embeddings to be conditioned on the phonetic content of the input utterances. Refer to \cite{peng2023MHFA} for more details.

\subsubsection{Reducing overfitting}

To mitigate the effect of overfitting from the WavLM font-end, these two aggregation methods rely on two components: (1) L2 regularization between the updated weights and the initial weights from the pre-trained WavLM model, which helps control overfitting caused by the large number of parameters; (2) layer-wise learning rate decay, following \cite{sun2020finetuneBERT}. Given the progressive abstraction of information across Transformer layers \cite{chen2022WavLM}, this technique allows more flexible weight updates in higher layers to adapt ASR capabilities, while ensuring lower layers preserve speech signals-related information.
\section{Experimental setup}
\label{sec:setup}

The front-end of all our models is based on the pre-trained WavLM Base model \footnote{\url{https://huggingface.co/microsoft/wavlm-base}}, it is composed of a CNN encoder and 12 Transformer layers. The dimension of each Transformer layer’s output is $768$. The number of parameters of the WavLM is $\sim 94M$, $1551$ for the linear weighted average pooling back-end, and $\sim 1M$ for the $32$ heads MHFA back-end.

All our models were trained on the whole training set released by ASVspoof 5 using an NVIDIA A100 80 GB GPU using the cross-entropy loss with a weight of $9$ for the bonafide class and $1$ for the spoof class to solve the class imbalance issue of the training set. The systems were trained on $4s$ speech utterance's length, randomly selected at each epoch from each training sample. We train for $100$ epochs with a default batch size of $120$ or $32$, and we stop the training if the EER on the scoring dataset does not improve after $50$ epochs. We use Adam optimizer with a learning rate of $5 \times 10^{-3}$ for the back-end and $2 \times 10^{-5}$ for the encoder, each reduced by $5\%$ every epoch. The test score was computed on the entire speech utterance. Results are reported in terms of Equal Error Rate (EER) and minimum Detection Cost Function (minDCF) following the setup described in \cite{ASVSpoof5evalplan}.

We train different models: first, by fixing the parameters of the encoders and training only the parameters of the MHFA back-end. Similar to previous work \cite{wang22_odyssey, tak22_odyssey}, we decide next to fine-tune the parameters of the encoder. We use a lower learning rate for the WavLM model to avoid overfitting. To make the training faster, these first experiments were conducted using only the noise and reverberation data augmentation. We also tested the efficiency of the RawBoost data-augmentation. Finally, we added the codec augmentation to the best configuration and retrained the model. 

\sisetup{
    output-decimal-marker = {.},
    detect-weight,
    mode=text
}
\newrobustcmd\B{\DeclareFontSeriesDefault[rm]{bf}{b}\bfseries}

\definecolor{Gray}{gray}{0.9}
\definecolor{Gray2}{gray}{0.8}
\begin{table*}[t]
  \caption{Spoof detection results of the different models trained during the ASVspoof 5 challenge on our scoring and progress datasets. The best performances are represented in bold text.}
  \label{tab:resultat}
  \centering
  \begin{footnotesize}
  \begin{tabular}{lccccccSSSS}
    \toprule    
    &\multicolumn{2}{c}{\textbf{Model}} & \textbf{Training} & \multicolumn{3}{c}{\textbf{Data-augmentation}} & \multicolumn{2}{c}{\textbf{Scoring Dataset}} & \multicolumn{2}{c}{\textbf{Progress Dataset}}\\
    \cmidrule(r){2-3} \cmidrule(r){4-4} \cmidrule(r){5-7} \cmidrule(r){8-9} \cmidrule(r){10-11}
                              
    \# & Back-end & Fine-tune WavLM & Batch size & Noise and RIR & Rawboost & Codec & \textbf{EER (\%)} & \textbf{$\text{minDCF}$} & \textbf{EER (\%)} & \textbf{$\text{minDCF}$} \\
    \midrule
    \rowcolor{Gray} & \multicolumn{2}{c}{Baseline (ResNet)} & 120 & $\checkmark$ &            & $\checkmark$ & 15.60 & 0.3469 & 16.19 & 0.3915 \\ 
    1 & MHFA  &                 & 120 &                &           &       & 6.78 & 0.1581 &    &     \\
    2 & MHFA  &                 & 120 & $\checkmark$   &           &       & 8.78 & 0.2155 &    &     \\
    3 & MHFA  & $\checkmark$    & 120 &               &           &       & 6.41 & 0.1628 &    &     \\ 
    4 & MHFA  & $\checkmark$    & 120 & $\checkmark$   &           &       & 3.37 & 0.0872 &   1.42 & 0.0380 \\
    5 & MHFA & $\checkmark$    & 120 &               &  $\checkmark$ &    & 28.91 & 0.7160 & & \\
    6 & MHFA & $\checkmark$    & 120 & $\checkmark$ &             & $\checkmark$ & 2.18 & 0.0552 & 1.22 & 0.0320 \\
    7 & MHFA & $\checkmark$    & 32 & $\checkmark$ &             & $\checkmark$ & \B 1.82 & \B 0.0498 & 1.13 & 0.0279 \\
    8 & WA & $\checkmark$    & 32 & $\checkmark$ &             & $\checkmark$ & 1.89 & 0.0503 & \B 1.01 & \B 0.0251 \\
    \rowcolor{Gray2} 9 &\multicolumn{6}{c}{\textbf{Fusion of model 6, 7 and 8}} & \B 1.10 & \B 0.0272 & \B 0.88 & \B 0.0226 \\ 
    \bottomrule
  \end{tabular}
\end{footnotesize}
\end{table*}

\section{Results and discussions}
\label{sec:result}

Our strategy consisted of the development of a main system that achieved the best possible individual performance before training a fusion that could improve the performance of the final system.

Table \ref{tab:resultat} summarizes some preliminary results obtained during the development of our main system. We report the performances on our scoring dataset and the progress dataset when the system was submitted during the progress phase of the challenge. We did not succeed in performing well with the baseline systems proposed in the ASVspoof~5 toolkit. We obtained the best performance for the baseline system based on a ResNet with noise, RIR, and codec augmentations.   

The results of the first four systems were expected: (1) the WavLM performed better than the baseline, this has been demonstrated on other speech processing tasks; (2) fine-tuning the WavLM is necessary to reach better performance. Data-augmentation is also fundamental as fine-tuning the WavLM with the noise and RIR augmentations allowed for reaching the best results. Since the WavLM has a large number of parameters, the model is more subject to overfitting in this case compared to the case where the WavLM weights are frozen.

Data augmentation becomes necessary to increase the performance of the WavLM and reduce overfitting; we experiment with different RawBoost algorithms proposed in \cite{tak2022rawboost}. 
We achieve an EER of $28.91\%$, which is worse than the performance obtained with the noise and RIR data-augmentation. 

With systems numbers 6 and 7, the performances obtained by using codec augmentation in addition to the noise and RIR augmentation are better. We obtain a $35\%$ relative improvement of the EER with a batch size of $120$ samples and a $47\%$ relative improvement of the EER with a batch size of $32$. This result was expected because it has been observed in practice that when using a larger batch, there is a degradation in the quality of the model, as measured by its ability to generalize \cite{LeCun2012, KeskarMNST17}. Initially, we selected a large batch size to make the experiments run faster using data parallelism.

To reduce the effect of overfitting, we implement a Weighted Average (WA) back-end, which has a limited number of parameters compared to the MHFA back-end. Thus, this model is less subject to overfitting compared to the MHFA. As expected, the result of this system was a little bit worse on our scoring dataset than the MHFA, but it performed better on the progress dataset and obtained an EER of $1.01$ and a minDCF of $0.0251$, which is our best performance on the progress dataset with an individual system. This result confirms that we would need more training samples or data augmentation algorithms to avoid overfitting when using the MHFA back-end.

In the end, fusion and calibration were performed using linear logistic regression with the Bosaris toolkit \cite{bosaris}. To select the best fusion combination, we implemented a greedy fusion scheme. First, we calibrated all the systems and selected the best, given the lowest minDCF cost. The best three systems were linearly fused to obtain the submission system. This fusion performed the best on the scoring and progress dataset. This result confirms the complementarity between the MHFA and WA back-ends.

\begin{table*}[!htb]
  \caption{Detailed performance of fusion system on the evaluation dataset according to different acoustic conditions.}
  \label{tab:resultatfusion}
  \centering
{\scriptsize
\begin{tabular}{cccccccccccccc}
\toprule
       &  pooled  &    -     & codec-1  & codec-10 & codec-11 & codec-2  & codec-3  & codec-4  & codec-5  & codec-6  & codec-7  & codec-8  & codec-9 \\ 
\midrule
pooled &  0.0937 &  0.0249 &  0.0644 &  0.1764 &  0.0726 &  0.0450 &  0.0728 &  0.1298 &  0.0311 &  0.0511 &  0.1824 &  0.1468 &  0.1096\\ 
 A17   &  0.0081 &  0.0000 &  0.0006 &  0.0076 &  0.0021 &  0.0004 &  0.0032 &  0.0053 &  0.0000 &  0.0007 &  0.0102 &  0.0256 &  0.0025\\ 
 A18   &  0.0328 &  0.0027 &  0.0144 &  0.0749 &  0.0118 &  0.0124 &  0.0186 &  0.0392 &  0.0069 &  0.0081 &  0.0693 &  0.0532 &  0.0388\\ 
 A19   &  0.1622 &  0.0624 &  0.1298 &  0.1633 &  0.0646 &  0.0830 &  0.1007 &  0.2096 &  0.0721 &  0.0987 &  0.2797 &  0.1539 &  0.1001\\ 
 A20   &  0.0631 &  0.0160 &  0.0386 &  0.0865 &  0.0169 &  0.0322 &  0.0242 &  0.0971 &  0.0229 &  0.0311 &  0.1432 &  0.0553 &  0.0427\\ 
 A21   &  0.0227 &  0.0013 &  0.0101 &  0.0340 &  0.0118 &  0.0045 &  0.0202 &  0.0261 &  0.0024 &  0.0042 &  0.0389 &  0.0555 &  0.0169\\ 
 A22   &  0.0506 &  0.0071 &  0.0279 &  0.1120 &  0.0225 &  0.0145 &  0.0518 &  0.0753 &  0.0079 &  0.0152 &  0.1080 &  0.0910 &  0.0489\\ 
 A23   &  0.0351 &  0.0033 &  0.0158 &  0.0735 &  0.0145 &  0.0107 &  0.0278 &  0.0467 &  0.0059 &  0.0098 &  0.0739 &  0.0552 &  0.0499\\ 
 A24   &  0.0968 &  0.0083 &  0.0604 &  0.2239 &  0.0711 &  0.0233 &  0.0980 &  0.1012 &  0.0107 &  0.0531 &  0.1388 &  0.2302 &  0.1327\\ 
 A25   &  0.0216 &  0.0020 &  0.0092 &  0.0364 &  0.0036 &  0.0062 &  0.0144 &  0.0303 &  0.0040 &  0.0034 &  0.0648 &  0.0372 &  0.0160\\ 
 A26   &  0.0261 &  0.0006 &  0.0102 &  0.0551 &  0.0130 &  0.0020 &  0.0226 &  0.0304 &  0.0002 &  0.0057 &  0.0533 &  0.0592 &  0.0297\\ 
 A27   &  0.0608 &  0.0074 &  0.0328 &  0.1427 &  0.0188 &  0.0275 &  0.0270 &  0.1155 &  0.0145 &  0.0197 &  0.1938 &  0.0830 &  0.0725\\ 
 A28   &  0.3332 &  0.0810 &  0.2403 &  0.6971 &  0.4099 &  0.1624 &  0.3527 &  0.3589 &  0.0890 &  0.2091 &  0.4293 &  0.7063 &  0.5634\\ 
 A29   &  0.0095 &  0.0011 &  0.0031 &  0.0095 &  0.0065 &  0.0012 &  0.0033 &  0.0054 &  0.0012 &  0.0024 &  0.0061 &  0.0308 &  0.0060\\ 
 A30   &  0.0641 &  0.0110 &  0.0357 &  0.1430 &  0.0203 &  0.0258 &  0.0317 &  0.1233 &  0.0174 &  0.0246 &  0.1923 &  0.0873 &  0.0612\\ 
 A31   &  0.1172 &  0.0283 &  0.0815 &  0.2383 &  0.0495 &  0.0511 &  0.0745 &  0.1831 &  0.0400 &  0.0598 &  0.2719 &  0.1740 &  0.1303\\ 
 A32   &  0.0504 &  0.0060 &  0.0271 &  0.1176 &  0.0093 &  0.0217 &  0.0164 &  0.1096 &  0.0113 &  0.0173 &  0.1843 &  0.0550 &  0.0440\\ 
\bottomrule
\end{tabular}
}

\centerline{(a) Minimum Detection Cost Function}

\vspace{9pt}

{\scriptsize
\begin{tabular}{cSSSSSSSSSSSSS}
\toprule
       &  \multicolumn{1}{c}{pooled}  &    \multicolumn{1}{c}{-}     & \multicolumn{1}{c}{codec-1}  & \multicolumn{1}{c}{codec-10} & \multicolumn{1}{c}{codec-11} & \multicolumn{1}{c}{codec-2}  & \multicolumn{1}{c}{codec-3}  & \multicolumn{1}{c}{codec-4}  & \multicolumn{1}{c}{codec-5}  & \multicolumn{1}{c}{codec-6}  & \multicolumn{1}{c}{codec-7}  & \multicolumn{1}{c}{codec-8}  & \multicolumn{1}{c}{codec-9} \\ 
\midrule
pooled &   3.42  &   0.92  &   2.49  &   6.45  &   3.16  &   1.66  &   3.00  &   4.66  &   1.13  &   2.04  &   6.37  &   6.02  &   4.45 \\ 
 A17   &   0.30  &   0.00  &   0.04  &   0.26  &   0.09  &   0.01  &   0.15  &   0.20  &   0.00  &   0.05  &   0.37  &   0.92  &   0.14 \\ 
 A18   &   1.17  &   0.10  &   0.50  &   2.67  &   0.46  &   0.46  &   0.71  &   1.44  &   0.29  &   0.39  &   2.43  &   1.92  &   1.41 \\ 
 A19   &   5.61  &   2.19  &   4.69  &   5.82  &   2.32  &   2.93  &   3.56  &   7.29  &   2.49  &   3.65  &   9.89  &   5.70  &   3.60 \\ 
 A20   &   2.18  &   0.55  &   1.37  &   3.01  &   0.59  &   1.14  &   0.86  &   3.36  &   0.81  &   1.12  &   5.11  &   2.03  &   1.54 \\ 
 A21   &   0.82 &   0.05  &   0.46  &   1.21  &   0.50  &   0.18  &   0.78  &   1.07  &   0.12  &   0.14  &   1.42  &   1.92  &   0.68 \\ 
 A22   &   1.79  &   0.28  &   1.09  &   3.88  &   0.80  &   0.54  &   1.94  &   2.77  &   0.29  &   0.53  &   3.85  &   3.18  &   1.74 \\ 
 A23   &   1.23  &   0.13  &   0.58  &   2.67  &   0.55  &   0.41  &   1.03  &   1.69  &   0.22  &   0.43  &   2.58  &   1.97  &   1.74 \\ 
 A24   &   3.44  &   0.32  &   2.22  &   8.43  &   2.53  &   0.96  &   3.64  &   3.61  &   0.41  &   1.94  &   4.98  &   8.13  &   4.75 \\ 
 A25   &   0.75  &   0.07  &   0.38  &   1.42  &   0.12  &   0.22  &   0.54  &   1.08  &   0.13  &   0.14  &   2.37  &   1.39  &   0.58 \\ 
 A26   &   0.94  &   0.02  &   0.42  &   1.92  &   0.51  &   0.08  &   0.86  &   1.12  &   0.01  &   0.20  &   2.01  &   2.05  &   1.06 \\ 
 A27   &   2.19  &   0.28  &   1.23  &   5.40  &   0.68  &   1.09  &   0.99  &   4.29  &   0.55  &   0.69  &   6.89  &   2.90  &   2.64 \\ 
 A28   &  12.01  &   3.03  &   8.61  &  25.29  &  15.13  &   6.30  &  12.75  &  12.85  &   3.33  &   8.02  &  15.31  &  24.79  &  21.20 \\ 
 A29   &   0.39  &   0.07  &   0.25  &   0.50  &   0.42  &   0.04  &   0.25  &   0.33  &   0.04  &   0.19  &   0.37  &   1.21  &   0.48 \\ 
 A30   &   2.28  &   0.41  &   1.26  &   5.04  &   0.75  &   1.00  &   1.12  &   4.32  &   0.63  &   0.87  &   6.69  &   3.04  &   2.19 \\ 
 A31   &   4.07  &   1.04  &   2.93  &   8.32  &   1.74  &   1.86  &   2.63  &   6.47  &   1.39  &   2.11  &   9.44  &   6.02  &   4.75 \\ 
 A32   &   1.84  &   0.22  &   1.02  &   4.30  &   0.34  &   0.89  &   0.67  &   4.07  &   0.39  &   0.63  &   6.64  &   2.04  &   1.68 \\ 
\bottomrule
\end{tabular}
}

\vspace{3pt}
\centerline{(b) EER}
\end{table*}

On the evaluation dataset, the fused system achieves $0.0937$ minDCF, $3.42\%$ EER, $0.1927$ Cllr, and $0.1375$ actDCF. Unlike our previous results with the progress dataset, this performance is worse than on our scoring dataset. This results from new acoustic conditions where the model could not generalize better.

We report in Table \ref{tab:resultatfusion} the detailed performances of our submitted system according to the different acoustic conditions. The first analysis of these two tables shows that condition A28, which uses audio speech generated using the pre-trained YourTTS model \cite{pmlr-v162-casanova22a}, is the most challenging task in our case. A detailed analysis shows that using limited bandwidth codec compression is also difficult because we lose speech information in higher frequencies. Finally, we can notice that some specific combinations are very challenging such as A28-codec10 and A28-codec8.
\section{Conclusions}
\label{sec:conclusion}

In this article, we have presented our countermeasure systems based on the pre-trained WavLM Base model for the ASVspoof 5 challenge open condition task. These systems significantly outperformed the baseline. We have shown that this model can be a good feature extractor for a back-end detection system. Similar to previous work based on large models such as wav2vec 2.0, this model needs to be fine-tuned using a spoofed dataset. The MHFA back-end obtained good performance on our development dataset, but it was more subject to overfitting than WA, which has fewer parameters. This simple Weighted Average (WA) pooling obtains the best performances on the progress dataset. We would need more training samples and augmentation algorithms to avoid this issue. The fusion of the systems based on MHFA and WA achieved the best performance and confirmed the complementary relationship between the two techniques.  As WavLM representations also contain valuable speaker identity information, we could explore combining the two tasks with a back-end for each downstream task.

\section{Acknowledgements}

This work was performed using HPC resources from GENCI-IDRIS (Grant 2023-AD011014623) and has been partially funded by the French National Research Agency (project APATE - ANR-22-CE39-0016-05).

\bibliographystyle{IEEEbib}
\bibliography{mybib}

\end{document}